\documentclass[twocolumn,prl,aps,superscriptaddress,showpacs,amsmath,amssymb,floatfix]{revtex4-1}
\usepackage{color}
\usepackage{xcolor}
\usepackage{mathrsfs}
\usepackage{amsmath,wasysym}
\usepackage{graphicx}
\usepackage{dcolumn}
\usepackage{bm}
\usepackage{times}
\usepackage{amssymb}
\usepackage{float}
\usepackage{array}
\usepackage{ulem}

\begin{document}

\title{Dynamical phases in quenched spin-orbit coupled degenerate Fermi gas}
\author{Ying Dong}
\affiliation{Department of Physics, Hangzhou Normal University, Hangzhou, Zhejiang 310036, China}
\author{Lin Dong}
\affiliation{Department of Physics and Astronomy, and Rice Quantum Institute, Rice University, Houston, Texas 77251-1892, USA}
\author{Ming Gong}
\affiliation{Department of Physics and Center of Coherence, The Chinese University of Hong Kong, Shatin, N.T., Hong Kong, China}
\author{Han Pu}
\affiliation{Department of Physics and Astronomy, and Rice Quantum Institute, Rice University, Houston, Texas 77251-1892, USA}

\date{\today}

\begin{abstract}
\textbf{The spin-orbit coupled degenerate Fermi gas provides a new platform to realize topological superfluids and related topological excitations. However previous studies are mainly focused on the topological properties of the stationary ground state. Here we investigate the quench dynamics of a spin-orbit coupled two-dimensional Fermi gas, in which the Zeeman field serves as the major quench parameter. Three post-quench dynamical phases are identified according to the asymptotic behavior of the order parameter. In the undamped phase, a persistent oscillation of the order parameter may support a topological Floquet state with multiple edge states. In the  damped phase, the magnitude of the order parameter approaches a constant via a power-law decay, which may support dynamical topological phase with one edge state at the boundary. In the over-damped phase, the order parameter decays to zero exponentially although the condensate fraction remains finite. These predictions can be observed in the strong coupling regime.}
\end{abstract}

\maketitle

Quantum statistical mechanics provides a general framework under which the
characterization of equilibrium states of many-body systems is well established.
To describe systems slightly perturbed out of equilibrium, the linear response
theory turns out to be extremely successful. Much is less known, however, on the
quantum coherent dynamics of a system far-from-equilibrium. A main reason for
that can be ascribed to the fact that such dynamics is generally difficult to access in experiments due to unavoidable relaxation and dissipation from interactions with the environment. However, the dynamics of quantum systems far-from-equilibrium is of great interest from a fundamental viewpoint because it can provide us with the properties of the system beyond ground state, for instance, excitations, thermalization, (dynamical) phase transitions and related universalities \cite{Endres12, Kinoshita06, Gring12, Sadler06, ZurekPT, Hoff}; see more details in a recent Review  \cite{Anatoli}. In this regard,  ultracold atom systems provide a new platform for the exploration of intriguing far-from-equilibrium {\it coherent} dynamics \cite{Sadler06,Sommer11,Chin10}. This is made possible by the precise control of key parameters in cold atomic systems as well as the ideal isolation from environment \cite{Lewenstein10}.

For these reasons, the coherent dynamics of the $s$-wave Bardeen-Cooper-Schrieffer (BCS) superfluid has been intensively studied over the past decade  \cite{Volkov, Barankov04,Andreev04,Yuzbashyan06a,Yuzbashyan06b, Barankov06,Bulgac09, Yuzbashyan05}. In these theoretical studies, via the so-called Anderson's pseudospin representation, the BCS model can be {\it exactly} mapped into a classical spin model, which is proven to be integrable and can be solved exactly using the auxiliary Lax vector method \cite{LaxV, LaxV2}. It has been shown that the quench dynamics of the system depends strongly on both the initial and the final values of the quench parameter, which is often chosen to be the interaction strength. In general, three different phases can be identified according to the long-time asymptotic behavior of the order parameter: the undamped oscillation phase (synchronization phase), the damped oscillation phase, and the over-damped phase. Integrability of the Hamiltonian is essential to understand these results. The $p$-wave superfluid has the same mathematical structure as the $s$-wave superfluid, thus similar phases are observed in a recent study  \cite{Foster13, Foster14}. As the $p$-wave superfluid supports topological phases, a quenched $p$-wave superfluid is found to support dynamical topological phases within certain parameter regimes. However, Fermi gas with $p$-wave interaction, realized by tuning the system close to a $p$-wave Feshbach resonance, suffers from strong incoherent losses  due to inelastic collisions \cite{Gaebler07}. On the other hand, one may realize cold atomic systems with effective $p$-wave interaction. Such candidates include ultracold polar molecules with long-range dipolar interaction\cite{Levinson} and spin-1/2 Fermi gases subject to synthetic partial waves\cite{Williams12}. Both of these systems are being intensively investigated in cold atom research.

In this Article, we study the quench dynamics and topological edge states in a
spin-orbit (SO) coupled superfluid Fermi gas in two dimension (2D), motivated by
the very recent realization of SO coupling in ultracold atoms
\cite{Lin11,Zhang12,Wang12,Cheuk12, Qu13, Chris14}. The ground state of this
system can be topologically nontrivial in some parameter regimes  \cite{MGong11,
MGong12, Sato09,ZhouJ11, Pu3, Pu5, Seo12,Iskin11,HuiH11,Pu2,Pu9,
CQu13,WZhang13,Liu13,CC13, shenoy1, shenoy2, shenoy3}. This is because the SO
coupling, Zeeman field and $s$-wave interaction together can lead to effective
$p$-wave pairing. This system possesses several control parameters that can be
readily tuned in experiments, which makes it ideal to study the
far-from-equilibrium {\it coherent} dynamics and related topological phase
transitions. However, the simultaneous presence of the SO coupling and the
Zeeman field breaks the integrability of this model\cite{JiaL}, and change the
system from a single-band to a two-band structure. It is therefore natural to
ask the fundamental question: "What types of post-quench dynamical phases this
system will exhibit, and how do these dynamical phases differ from the ones
supported by the integrable models?" In our study, we choose the Zeeman field as
the quench parameter for the reasons to be discussed below. It is quite
surprising that all the phases supported by the integrable model still exist in
our non-integrable system, although there exists unique topological features in
our system. We provide a complete phase diagram and investigate each phases in
details. We also show how dynamical topological phases, which can support topologically protected edge states, emerge in
this model. \newline

{\Large Results} \newline

{\bf Model Hamiltonian}
We consider a 2D system of uniform  SO coupled degenerate Fermi gas with $s$-wave interaction confined in the $xy$-plane, whose
Hamiltonian is written as $\mathcal{H} = \mathcal{H}_0+\mathcal{V}$, where ($\hbar = 1$)
\begin{eqnarray}
\mathcal{H}_0 & =& \sum_{{\bf k},s,s'}c_{{\bf k}s}^\dag\left[\xi_{\bf k}+\alpha(k_y\sigma_x-k_x\sigma_y)+h\sigma_z\right]_{ss'}c_{{\bf k}s'}, \nonumber \\
\mathcal{V}   & =& g\sum_{{\bf k}, {\bf k}', {\bf q}} c_{{\bf k+q}\uparrow}^\dagger c_{{\bf k}'-{\bf q}\downarrow}^\dagger c_{{\bf k}'\downarrow} c_{{\bf k}\uparrow},
\label{eq-H1}
\end{eqnarray}
where $s, s' = \uparrow,\downarrow$ label the pseudospins represented by two atomic hyperfine states, ${\bf k}$ is the momentum operator,
$\xi_{\bf k}=\epsilon_{\bf k}-\mu$, where $\epsilon_{\bf k}={\bf k}^2/2m$ denotes  kinetic energy and $\mu$ is the chemical potential,
$\alpha$ is the Rashba SO coupling strength, $\sigma_{x,y,z}$ are the Pauli matrices, $h$ is the Zeeman field along the $z$-axis, 
$c_{{\bf k}s}$ is the annihilation operator which annihilates a fermion with momentum ${\bf k}$ and spin $s$, and $g$ represents the inter-species $s$-wave interaction strength. Notice that the 2D system is
created from the three dimensional system by applying a strong confinement along the $z$-axis, thus $g$ in principle can be controlled by both confinement and Feshbach resonance. This 2D degenerate Fermi gas has been realized in recent experiments \cite{Martiyanov10, Martiyanov14, Frohlich11, Dyke2011}.

We consider the quench dynamics of the Fermi gas at the mean field level, thus the potential defect production, i.e., the so-called Kibble-Zurek mechanism, after the quench is not considered. Imagine that we prepare the initial system in the ground state with Zeeman field $h_i$. At time $t =0^+$, we suddenly change the Zeeman field from $h_i$ to some final value $h_f$. This scheme should be in stark contrast to previous literatures, in which the interaction strength $g$ generally serves as the quench parameter \cite{Volkov, Barankov04, Andreev04,Yuzbashyan06a,Yuzbashyan06b, Barankov06,Bulgac09, Foster13, Yuzbashyan05, Foster14}. We choose the Zeeman field as our quench parameter for the following reasons. (1) As we shall show in the discussion of the ground state properties of this system, the Zeeman field directly determines the topological structure of the ground state. (2) In SO coupled quantum gases, the laser intensity and/or detuning serve as effective Zeeman field, and these parameters can be changed in a very short time scale, satisfying the criterion for a sudden quench. By contrast, a change of the interaction strength is achieved by tuning the magnetic field, via Feshbach resonance, which usually cannot be done very rapidly. (3) Moreover, changing the effective Zeeman field in SO coupled quantum gases has already been demonstrated in recent experiments \cite{Lin11,Zhang12,Wang12,Cheuk12, Qu13}, in which both the magnitude and the sign of the Zeeman field can be changed.

The superfluid order parameter
is defined as $\Delta = g\Sigma_{\bf k}\langle c_{-{\bf k}\downarrow}c_{{\bf k}\uparrow}\rangle$ and the interaction term can be rewritten as $\mathcal{V} \rightarrow \Delta^* c_{-{\bf k}\downarrow}c_{{\bf k}\uparrow}+\Delta c_{{\bf k}\uparrow}^\dag c_{-{\bf k}\downarrow}^\dag-|\Delta|^2/g$. Therefore, under the Numbu spinor basis $\Psi_{\bf k}=(c_{{\bf k}\uparrow},c_{{\bf k}\downarrow},-c_{-{\bf k}\downarrow}^\dag,c_{-{\bf k}\uparrow}^\dag)^T$, the total Hamiltonian becomes
$\mathcal{H}=\Sigma_{{\bf k}}\Psi^\dag_{\bf k} \mathcal{M}_{\bf
k}\Psi_{\bf k}-|\Delta|^2/g$,
where the Bogoliubov-de Gennes (BdG) operator reads as  \cite{MGong11, MGong12}
\begin{equation}
\mathcal{M}_{\bf k}=
{1 \over 2}
\begin{pmatrix}
H_0({\bf k})  & \Delta  \\
\Delta^* & -\sigma_y H_0^*(-{\bf k})\sigma_y
\end{pmatrix}.
\label{eq-Heff}
\end{equation}
Assuming $f_{{\bf k}\pm}=(u_{{\bf k}\pm},v_{{\bf k}\pm},p_{{\bf k}\pm},q_{{\bf k}\pm})^T$ are the two energy levels of $\mathcal{M}_{{\bf k}}$ with
positive eigenvalues, the order parameter can be determined by solving the gap equation $\Delta=-g\Sigma_{\bf k}(v_{{\bf k}+}q^*_{{\bf k}+}+v_{{\bf k}-}q^*_{{\bf k}-})$ self-consistently. Here the bare interaction strength $g$ should be regularized by $1/g = -\sum_{{\bf k}} 1/(2\epsilon_{{\bf k}} + E_b)$  \cite{MGong12}.
As a result, in the following, the interaction strength is quantitatively defined by the binding energy $E_b \in [0, \infty)$.

The coherent dynamics of this model cannot be solved due to the lack of any nontrivial symmetry \cite{JiaL}, thus it should be computed numerically.
As a reasonable but general assumption, we assume that the wavefunction, after quench,
is still BCS-like, i.e., $|\psi(t)\rangle = \prod_{{\bf k}, s = \pm} f_{{\bf k}\pm}^\dagger(t) \Psi_{{\bf k}}|0\rangle$,
where the dynamics of the vectors $f_{{\bf k}\pm}(t)$ are determined by the following
time-dependent BdG equation
\begin{equation}
\label{eq-tBdG}
i{\partial \over \partial t} f_{{\bf k}\pm} = \mathcal{M}_{{\bf k}} f_{{\bf k}\pm}.
\end{equation}
Here $\mathcal{M}_{\bf k}$ is the time-dependent BdG Hamiltonian in which $\Delta(t)$ now evolves in time after the quench.

\begin{figure}
\includegraphics[width=3.4in]{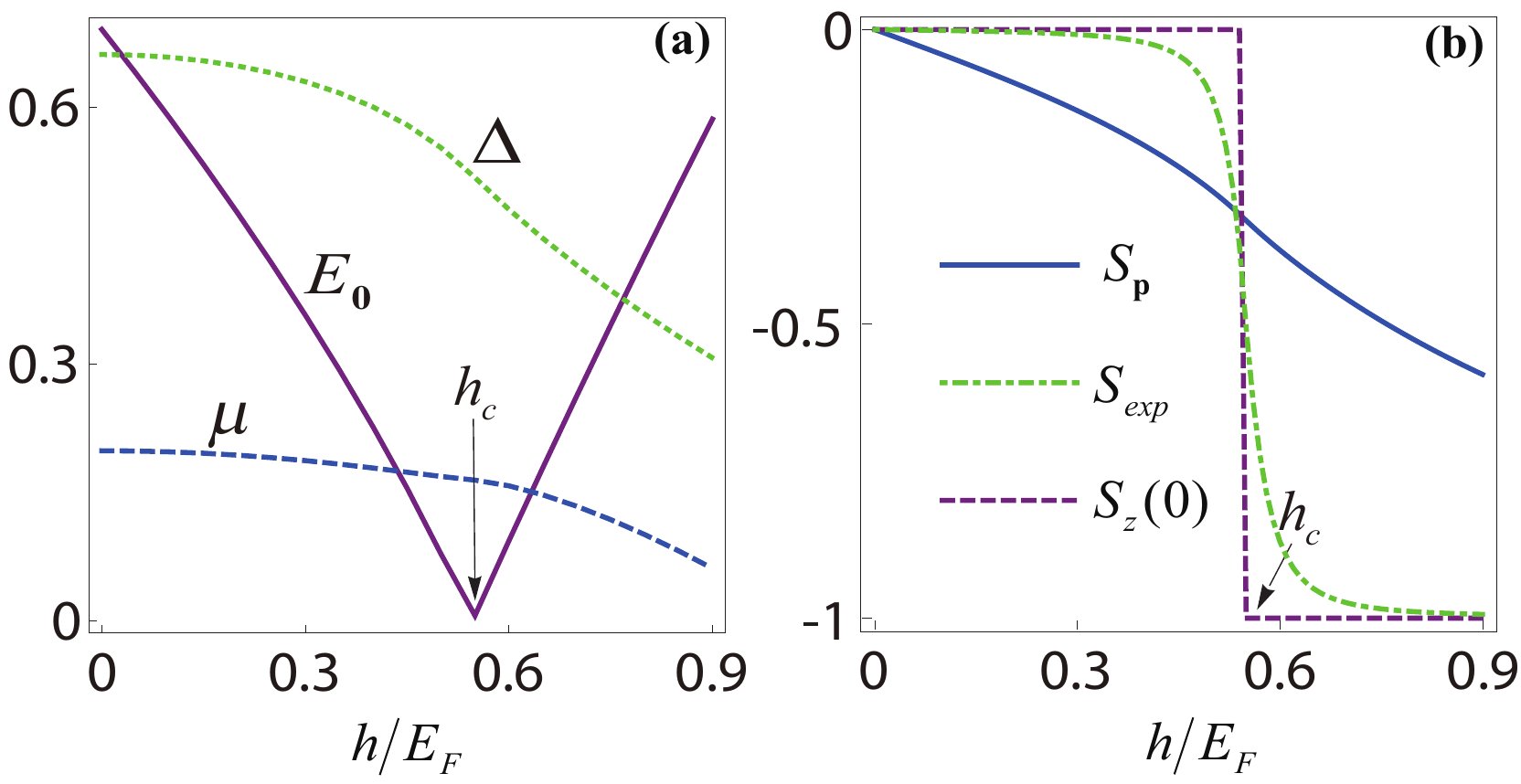}
\caption{{\bf Topological phase transition in spin-orbit coupled superfluids}. (a) Plot of $E_0$ (energy gap at zero momentum), $\Delta$, and $\mu$ as functions of $h$. The arrow marks the critical Zeeman field $h_c\sim0.545E_F$. The topological phase transition is characterized by the topology of the ground state instead of symmetry because no spontaneous symmetry breaking takes place across the phase transition point. (b) Plot of the total spin polarization $S_\text{p}$ and zero momentum spin population $S_z(0)$ as a function of Zeeman field. $S_\text{p}$ is a smooth function of $h$, whereas $S_z(0)$ jumps at $h_c$ due to band inversion transition. $S_\text{exp}$ is the spin population averaged over a region in momentum space (see text), which mimics the possible finite momentum resolution in realistic experiments. Throughout this paper if not specified otherwise we take $E_b=0.2E_F$ and $\alpha k_F=1.2E_F$.}
\label{fig-fig1}
\end{figure}

\begin{figure*}
\includegraphics[width=0.99\textwidth]{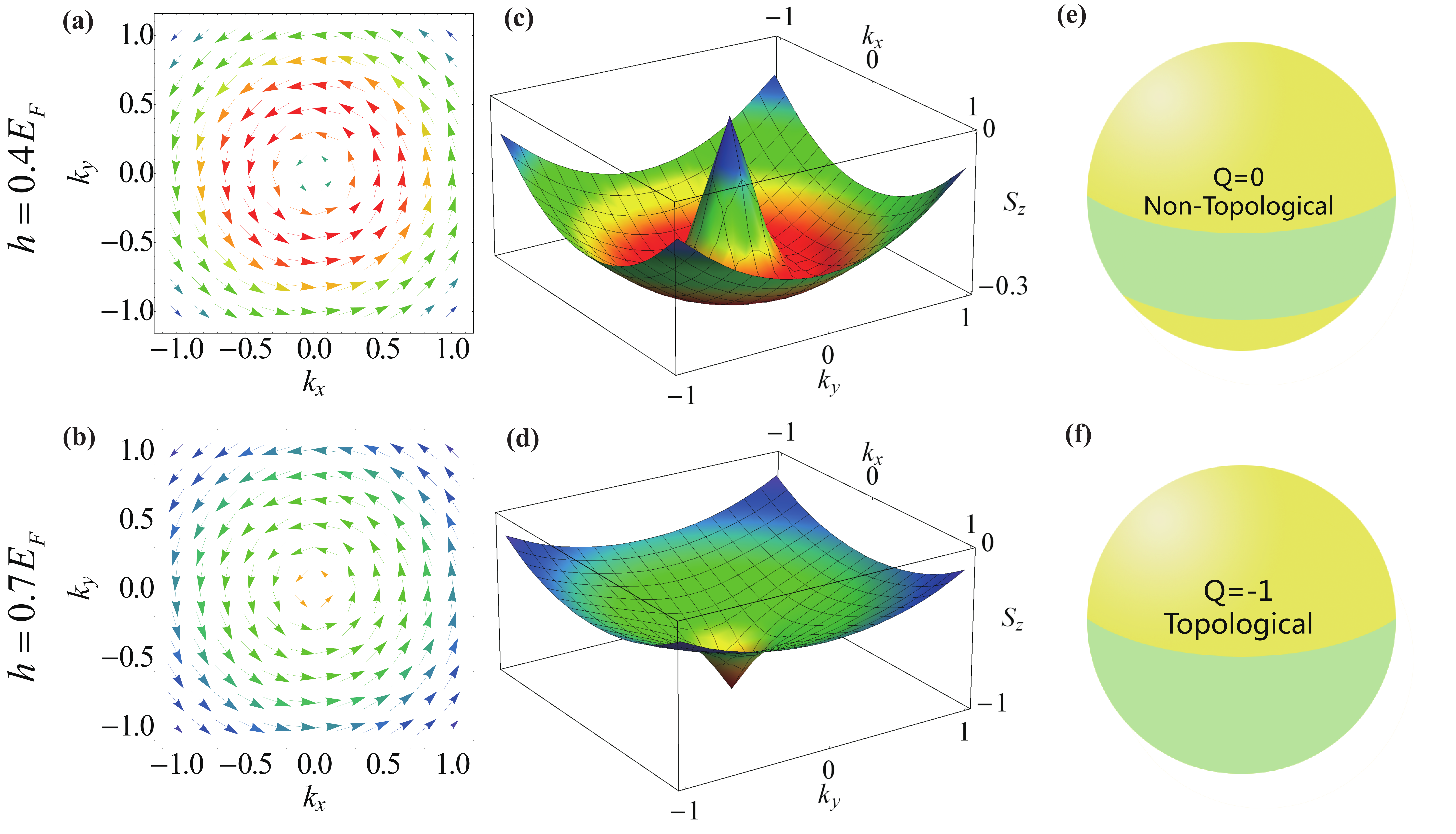}
\caption{ {\bf Topological properties of the spin texture}. (a), (b) Spin texture for the normalized spin vector ${\bf s}$ (Skyrmion) of the ground state of a non-topological state ($h = 0.4E_F$) and a topological state ($h = 0.7E_F$), respectively. For the parameter we used, $h_c \sim 0.545E_F$, see also Fig.~\ref{fig-fig1} (a). The corresponding $z$-component of the spin vector, $S_z({\bf k})$, are plotted in (c) and (d). (e) and (f) represent schematic plot of
the mapping from the 2D momentum space onto a unit sphere for non-topological phase and
topological phase, respectively. The darker green region represents the area swept out by the spin vector when the momentum ${\bf k}$ varies from 0 to $\infty$.}
\label{fig-fig2}
\end{figure*}

{\bf Ground State Properties}. Before we turn to the discussion of the quench dynamics, let us first briefly outline the ground state properties of the system. The most salient feature of the system is a topological phase transition induced by the Zeeman field  \cite{MGong11, MGong12, Sato09,ZhouJ11,Seo12,Iskin11,HuiH11,Pu2, Pu9, CQu13,WZhang13,Liu13,CC13}. Without the loss of generality, we assume that $h \ge 0$, in which case, the spin-up atom (spin-down atom) represents the minority (majority) species. It is well known that the topology of the superfluid is encoded in the topological index $W$, which corresponds to the topological state with $W=1$ if $h>\sqrt{\mu^2+\Delta^2}$, while $h<\sqrt{\mu^2+\Delta^2}$ yields $W=0$ and the state is non-topological. In Fig.~\ref{fig-fig1}(a), we show how $\mu$, $\Delta$ and the single-particle excitation gap $E_0$ at zero momentum change as a function of the Zeeman field $h$. At the critical point, $h_c=\sqrt{\mu^2+\Delta^2}$, the excitation gap $E_0$ vanishes, indicating a topological phase transition.  This feature is also essential for the realization of the long-time far-from-equilibrium {\it coherent} evolution; see Discussion.

To see this phase transition more clearly, we also examine the momentum space
spin texture. The spin vector is defined as ${\bf S}({\bf k})= ( c^\dag_{{\bf
k}\uparrow}, c^\dag_{{\bf k}\downarrow} ) {\boldsymbol \sigma} \left( c_{{\bf
k}\uparrow}, c_{{\bf k}\downarrow} \right)^T$, with the corresponding ground
state expectation value given by $\langle S_x({\bf k})\rangle=2\sum_{\eta=\pm}
\Re[p^*_{{\bf k}\eta}q_{{\bf k}\eta}]$, $\langle S_y({\bf
k})\rangle=2\sum_{\eta=\pm} \Im [p^*_{{\bf k}\eta}q_{{\bf k}\eta}]$, and
$\langle S_z({\bf k})\rangle=\sum_{\eta=\pm} (|q_{{\bf k}\eta}|^2-|p_{{\bf
k}\eta}|^2)$. As shown in Fig.~\ref{fig-fig1}(b), we find that $\langle S_z(0)
\rangle$, the spin component along the $z$-axis (which is just the population
difference between the two spin species) at ${\bf k}=0$, jumps discontinuously
when the Zeeman field crosses the critical value $h_c$, while the total spin
polarization which is defined as $S_\text{p} = (n_{\uparrow} -
n_{\downarrow})/n$, changes smoothly with respect to $h$. We emphasize that the
jump of $S_z(0)$ across the topological boundary is still visible if we take the
finite momentum resolution into account, as will be the case in realistic 
experiments; see curve 
$S_\text{exp}=\int d{\bf k}\, S_z({\bf k})/\int d{\bf k}\, n({\bf k})$ 
in Fig.~\ref{fig-fig1} (b), where $n({\bf k})$ is the total density at momentum ${\bf k}$ and the 
integral is performed in a circular area in momentum space centered at ${\bf k}={\bf 0}$ with a 
radius of $0.1k_F$. In fact, as shown below, the jump of $\langle S_z(0) \rangle$ just implies a change of topology of the spin texture.

Typical spin textures for $h<h_c$ and $h>h_c$ are plotted in Fig.~\ref{fig-fig2}(a) and (b), respectively. The topology of the system can be characterized by the Skyrmion number defined as
\begin{equation}
Q = \int\frac{d^2{\bf k}}{2\pi}\left[{\bf s}\cdot(\partial_{k_x}{\bf s}\times\partial_{k_y}{\bf s})\right],
\end{equation}
where ${\bf s}({\bf k}) =\langle {\bf S}({\bf k})\rangle/|\langle {\bf S}({\bf k})\rangle|$ is the normalized spin vector,
which maps the 2D momentum space to a unit sphere {\bf $S^2$}. In this sense, $Q$ is nothing but the number of times the spin vector wraps around the south hemisphere. Note that for momenta with fixed magnitude $|{\bf k}|$, $s_x$ and $s_y$ always sweep out a circle parallel to the equator. In the topologically trivial regime, we always have $s_z(0)=0$, see Fig.~\ref{fig-fig2}(c). Hence as $|{\bf k}|$ increases from zero, ${\bf s}$ begins at the equator, descends toward the south pole and then returns to the equator as $|{\bf k}|\rightarrow\infty$. Thus in this regime, ${\bf s}({\bf k})$ initially sweeps out the darker green region in the southern hemisphere of Fig.~\ref{fig-fig2}(e), but then unsweeps the same area, resulting in a vanishing winding number $Q=0$. In contrast, in the topologically nontrivial regime, we have $s_z(0)=-1$,  see Fig.~\ref{fig-fig2}(d). Hence as $|{\bf k}|$ increases from zero to infinity, the spin vector covers the entire southern hemisphere exactly once, as shown schematically in
Fig.~\ref{fig-fig2}(f), which leads to a nontrivial winding number $Q=-1$.  The sudden change of spin polarization is due to band inversion transition across the critical point.

It is worth pointing out that, for any quench, we always have $\frac{\partial}{\partial t}\langle S_z(0,t)\rangle=0$ (see Methods), regardless of the initial and final values of $h$. Therefore, $Q$ is unchanged over time after the quench. We can see that, the $Q$ and $W$ are equivalent at equilibrium, but their dynamics after a sudden quench are different. As shown below, $W$, which describes the topology of the full spectrum of the Hamiltonian, will evolve in time, while $Q$, which reflects the topological nature of the the state itself, will not. A similar conclusion is found in the study of the quenched $p$-wave superfluid \cite{Foster13, Foster14}. We emphasize that the momentum space spin texture studied here can be measured in cold atom experiments using the standard time-of-flight technique  \cite{Alba11,Cheuk12, Sau11}.

\begin{figure}
\label{phase}
\includegraphics[width=3.4in]{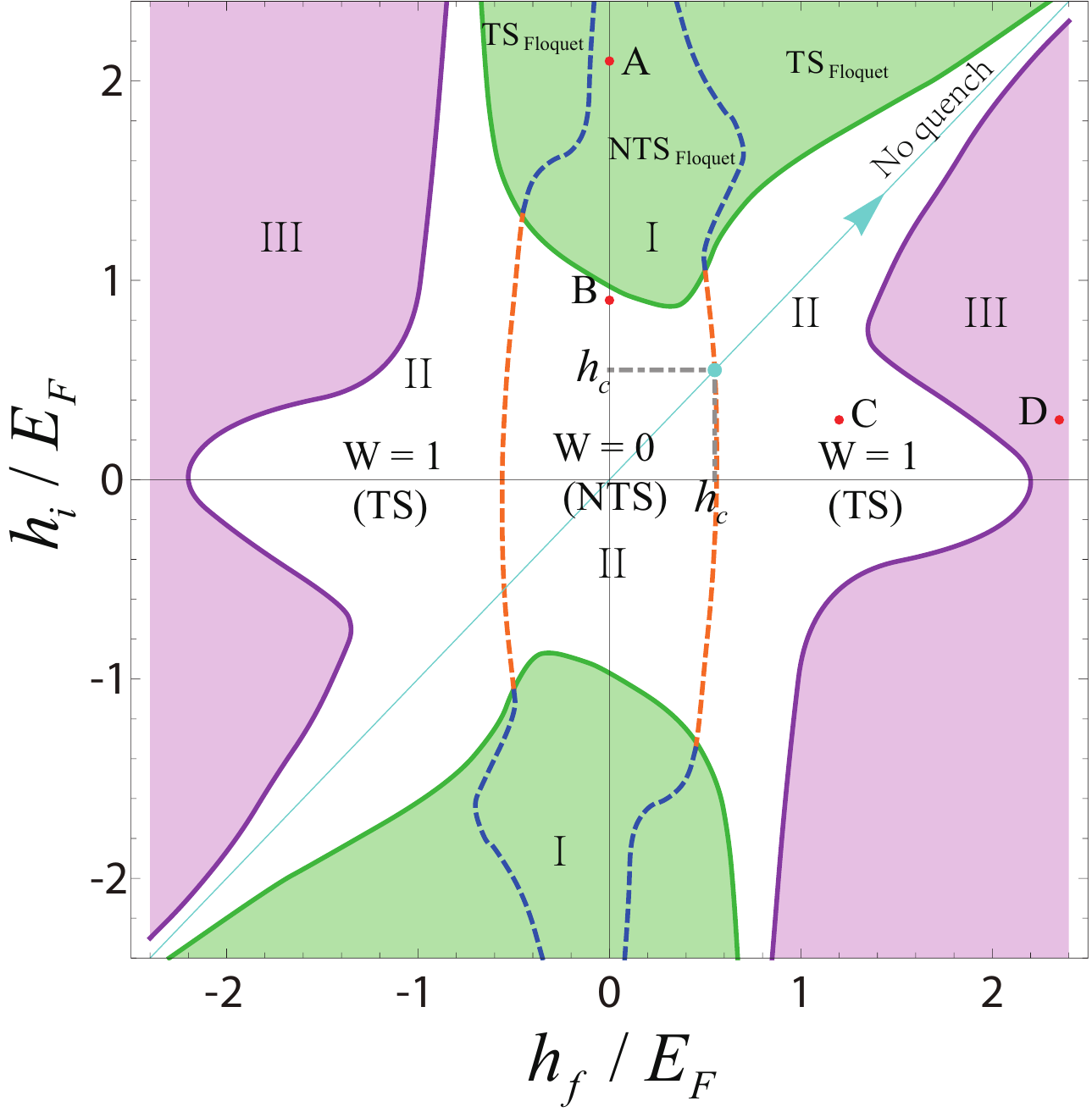}
\caption{{\bf Phase diagram of the quenched spin-orbit coupled superfluid condensate}. The different phases in this figure is obtained by the long-time
asymptotic behavior of the order parameter upon the quench of the Zeeman field
from initial value $h_i$ to final value $h_f$. The diagonal light blue line,
with $h_i=h_f$, is the case without quench, thus the quantum state is unchanged.
$h_c$ marks the quantum critical point separating the topological superfluid and
non-topological superfluid in the equilibrium ground state, which is determined
by $h_c^2 = \Delta^2 + \mu^2$. Three different dynamical phases observed in this
system are labeled with I, II, and III by green, white and purple shaded areas,
respectively. In phase I, $|\Delta(t)|$ shows persistent oscillations, which is
from the collisionless coherent dynamics. The dark blue dashed line separates
phase I into non-topological Floquet state denoted as
$\text{NTS}_\text{Floquet}$ and topological Floquet state labeled as
$\text{TS}_\text{Floquet}$. In phase II, $|\Delta(t)|\rightarrow\Delta_\infty$,
a nonzero constant value, which serves as the basic parameter to determine the
long-time asymptotic behavior.  The orange dashed lines are the non-equilibrium
extension of the topological phase transition at $h=h_c$, which separates phase
II into two parts, NTS and TS accordingly. Inside NTS (TS) region, the
energy spectrum of quasi-stationary Hamiltonian is trivial (nontrivial) without (with)
topologically protected edge modes. $W=0$ or $1$ marks the topological index at
$t = +\infty$.  In phase III, $|\Delta(t)|\rightarrow 0$ due to strong dephasing
from the out-of-phase collisions. We need to emphasize that the transition between various phases is smooth. However, these
different phases are well distinguished not very close to the boundaries (distance $>$
0.1 $E_F$). All other parameters are identical to Fig.~\ref{fig-fig1}. We have repeated the same calculation for several
different values of SO coupling strengths and found no qualitative differences.}
\label{fig-fig3}
\end{figure}

{\bf Dynamical phase diagram}. We now turn to our discussion on the quench dynamics. As in Refs. \cite{Foster13, Foster14}, we capture the dynamics using a phase diagram presented in Fig.~\ref{fig-fig3}. The phase diagram contains three different dynamical phases (which should not be confused with the equilibrium phases of a many-body system), identified by the distinct long-time asymptotic behavior of the order parameter in the parameter space spanned by the initial and final values of the Zeeman field $h_i$ and $h_f$. The three phases are labeled as phase I, II and III in Fig.~\ref{fig-fig3}. More specifically, in the undamped oscillation phase (phase I), the magnitude of the order parameter oscillates periodically without damping, although the wavefunction does not recover itself periodically. Away from the phase boundaries, the oscillation amplitude can be as large as a significant fraction of $E_F$ (see Fig.~\ref{fig-fig4} below). In the damped oscillation (phase II) regime, the order parameter exhibits damped oscillation with a power-law decay to a finite value. In the overdamped phase (phase III), the order parameter decays to zero exponentially. We also show that within  phase I and II, there exists dynamical topological regimes where topological edge states emerge in the asymptotic limit. Next, we shall discuss the properties of each phase in more details.

\begin{figure}
\includegraphics[width=3.4in]{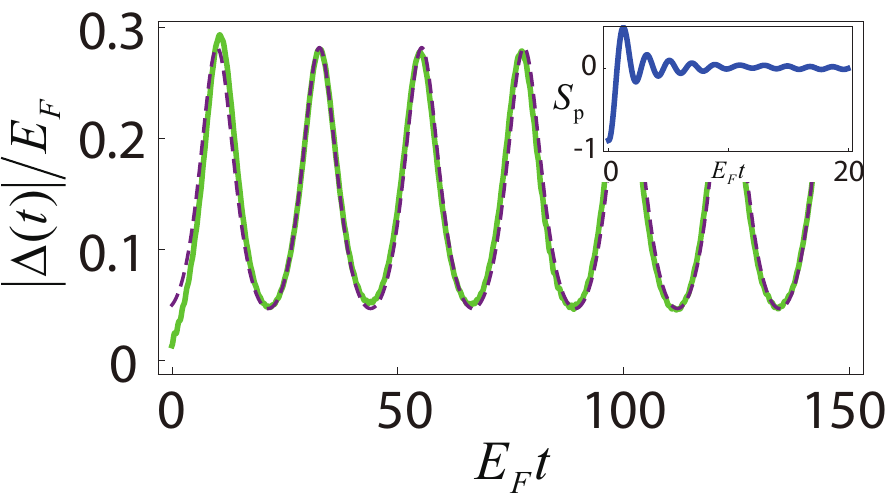}
\caption{{\bf Dynamics of order parameter and spin polarization in phase I}. We plot the result for point A $(h_i=2.1, h_f=0)E_F$ in Fig.~\ref{fig-fig3} with green solid line. The purple dashed line is the best fit using Eq. (\ref{elliptic}) with fitted parameter $\Delta_+=0.282E_F$, $\Delta_-=0.047E_F$. Inset shows the dynamics of spin polarization after quench, which quickly approaches a constant at the time scale of few $1/E_F$. }
\label{fig-fig4}
\end{figure}

{\bf Phase I.} In Fig.~\ref{fig-fig4} we plot the dynamics of the magnitude of the order parameter for a typical point in phase I (point A in Fig.~\ref{fig-fig3}), from which we find that $|\Delta(t)|$ oscillates asymptotically as  \cite{Barankov04}
\begin{equation}
\label{elliptic}
|\Delta(t)|=\Delta_+ \text{dn}[\Delta_+(t-\tau_0/2),\kappa], \quad   \kappa= 1-\Delta_-^2/\Delta_+^2,
\end{equation}
where $\text{dn}[u,k]$ is the periodic Jacobi elliptic function, and $\Delta_+$
and $\Delta_-$ are the maximum and minimum value of $|\Delta(t)|$, respectively.
This expression was first derived by Barankov {\it et al.} for the
weakly-interacting conventional BCS superfluid \cite{Barankov04}. The fitted
result using this empirical formula are presented in Fig.~\ref{fig-fig4} as dashed line, 
which fits the numerical results surprisingly well.  This phase can be 
understood from the picture of synchronization effect. In the previous studies of integrable BCS
models \cite{Barankov04, Barankov06, Foster14}, each Cooper pair---consisting of
two particles with momentum ${\bf k}$ and $-{\bf k}$---can be mapped into a
classical spin parameterized by ${\bf k}$, whose dynamics can be described as a
precession around an effective Zeeman field. The effective Zeeman field are
contructed from the order parameter and the kinetic energy of the Cooper pair.
For a large order parameter such that the contribution from the kinetic energy
is small, the effective Zeeman field is essentially the same for different
Cooper pairs and as a result, all Cooper pairs precess with roughly the same
frequency and phase coherence is therefore maintained. Here we reinterpret the
synchronization effect as a consequence of the strong condensation formed by Bose condensed Cooper pairs. 
The condensate ensure phase coherence of the constituent Cooper pairs. This reinterpretation is fully consistent with the spin precession picture, however it now becomes clear that the synchronization effect does not rely on whether the system is integrable or not. To substantiate this reinterpretation, we first notice that the Phase I region occurs in the parameter space where $|h_f|<|h_i|$, for which shortly after the quench the order parameter is increased. Second, we artificially include a finite temperature $T$ in the post-quench dynamical evoltion. For small $T$, the behavior of the order parameter remains essentially the same as the zero-temperature result. For large $T$, however, the order parameter no longer exhibits undamped oscillation, but rather decays to a constant. For the parameters used in Fig~\ref{fig-fig4}, this occurs when $T \apprge 0.2E_F$. Such a behavior can be easily undersood as a sufficiently large $T$ destroys the condensate and hence phase coherence between different Cooper pairs is lost.

To gain more insights into this persistent oscillating behavior, it is helpful
to investigate the spin population dynamics. We found that, as shown in the
inset of Fig.~\ref{fig-fig4}, the persistent oscillation of the order parameter
is not accompanied by a similar oscillation in the spin population. In fact, after the quench $S_{\rm p}$ exhibits damped oscillation and quickly reaches a steady-state value. Recently, the dynamics of the spin polarization after quench in a Fermi gas above the critical temperature has been measured in experiments  \cite{Wang12}. The decay of the spin polarization can be attributed to the interband Rabi oscillation, in which, different momentum state has slightly different Rabi frequency, such that destructive interference gives rise to the  damping phenomenon.

\begin{figure}
\centering
\includegraphics[width=3.4in]{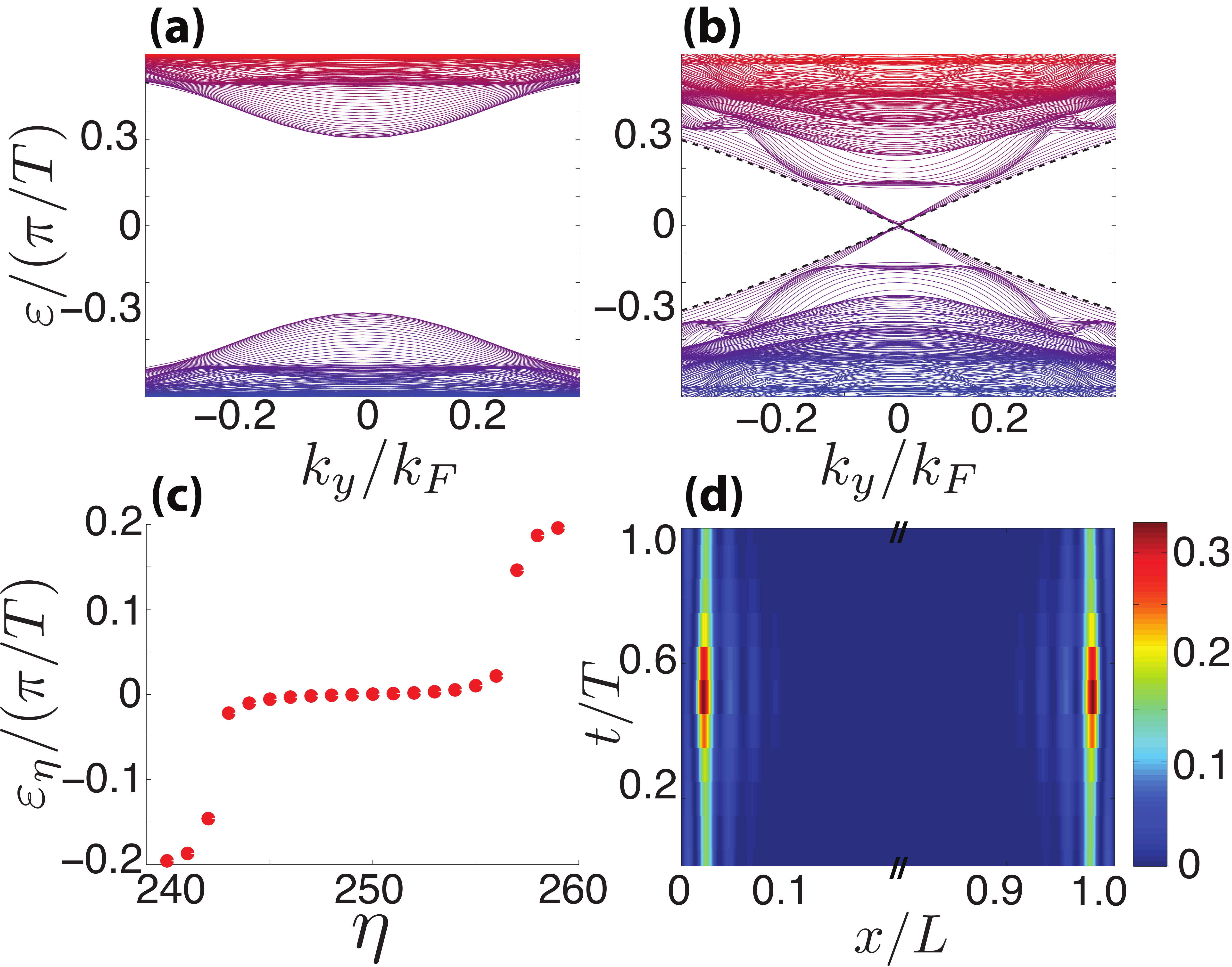}
\caption{{\bf Dynamical Floquet state in a strip for phase I}.  Imposing a strip geometry, we find the quasiparticle spectrum is trivial (gapped) in (a)
$(h_i=1.5, h_f=0.3)E_F$ and topologically nontrivial (gapless) in (b) $(h_i=2.1, h_f=0.9)E_F$. Seven pairs of edge states with linear dispersion at small ${\bf k}$ have been identified in (b), which is a direct manifestation of bulk topology based on bulk-edge correspondence. (c) An enlarged view of the low-lying quasiparticle spectrum at $k_y = 0$. Here $n$ is the index for quasiparticle states (see Methods). (d) Evolution of the wavefunction for one pair of edge states (the two symbols shown in purple in (c), $n=249$ and $n=250$) in one full period. Here we only plot the $|u|^2$ component of the wavefunction, and the other components show a similar behavior, i.e., they are also well localized near the boundary.}
\label{fig-fig5}
\end{figure}

Equation (\ref{elliptic}) provides an empirical formula for the asymptotic dynamics of the $|\Delta(t)|$. The order parameter itself behaves like
 $\Delta(t) = |\Delta(t)|e^{-i2\mu_{\infty}t +i \varphi(t)}$ \cite{Foster13, Foster14}, where $\varphi(t)$ (modulo $2\pi$) is also a periodic function with commensurate period to $|\Delta(t)|$. The phase factor's linear piece in time, $-2\mu_\infty t$, can be gauged out by a unitary transformation (see Methods) \cite{Foster14}. After the gauge transformation, we obtain a BdG Hamiltonian $\tilde{\mathcal{M}}_{\bf k}(t)$ that is periodic in time. For such a periodic system, we may invoke the Floquet theorem to examine its Floquet spectrum. To determine whether the system is topological in the Floquet sense, we calculate the spectrum $\varepsilon_{\text{\bf k}\pm}$ in a strip by adding a hard-wall boundary condition in the $x$-direction. Two examples of the spectrum are plotted in Fig.~\ref{fig-fig5}. In the example shown in Fig.~\ref{fig-fig5}(a), the spectrum is gapped, corresponding to a topologically trivial Floquet state. By contrast, the spectrum shown in Fig.~\ref{fig-fig5}(b), exhibits gapless modes at $k_y=0$. An enlarged view of these modes are presented in Fig.~\ref{fig-fig5}(c). Note that due to the finite-size effect, the energies of these gapless modes are not exactly zero, but on the order of $10^{-3}E_F$ and are well separated from the rest of the spectrum. In Fig.~\ref{fig-fig5}(d), we show the evolution of the wavefunctions of one pair of the gapless modes over one oscillation period. As one can see, the gapless modes are well localized at the boundaries. Furthermore, we have verified that their wavefunctions satisfy the requirement for Majorana modes. Hence these gapless excitations represent the Majorana edge modes and the system can therefore be characterized as a topological Floquet state. We have also verified that the edge states with the same chirality are localized at the same boundary. Thus no direct coupling is allowed between them due to particle-hole symmetry (see Method). This explains the robustness of the edge modes in a perturbative viewpoint, even though there can exist multiple edge modes near the same boundary. For more details, see Methods. We note that the number of the edge modes depends on the parameter set. For the example shown in Fig.~\ref{fig-fig5}(b), there are 7 pair of edge modes. In the phase diagram of Fig.~\ref{fig-fig3}, inside phase I, the two dark blue dashed lines characterize the topological boundaries, which separate the non-topological states denoted by $\text{NTS}_\text{Floquet}$ from the topological states denoted by $\text{TS}_\text{Floquet}$.

\begin{figure}
\includegraphics[width=3.4in]{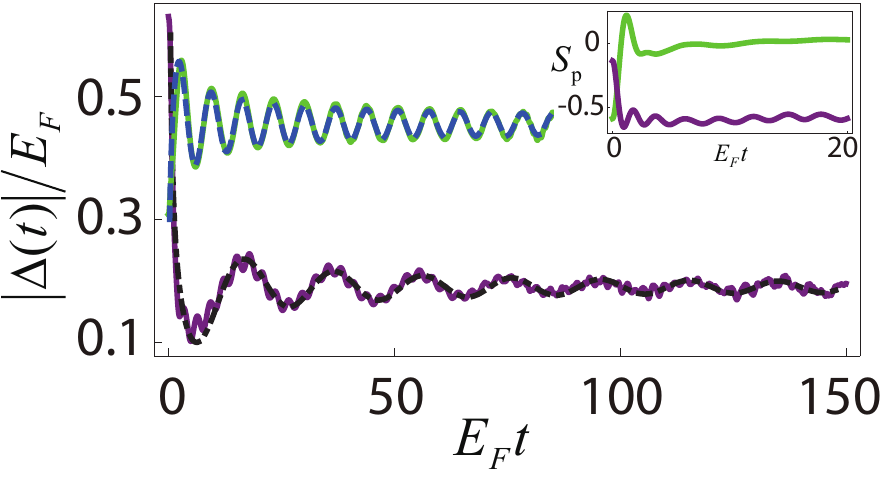}
\caption{{\bf Dynamics of order parameter and spin texture in phase II}. Green line represents dynamics for point B $(h_i=0.9, h_f=0)E_F$ and
the purple line is the result for point C $(h_i=0.3, h_f=1.2)E_F$ in Fig.~\ref{fig-fig3}, and dashed lines represent the fitted curve using Eq.~(\ref{cos}).
For point B (blue dashed line), $\Delta_{\infty}=0.455E_F$,   $E_\text{g}^\infty=0.456E_F$, $A=-0.16 E_F^{1+\nu}$, $\theta =\pi/4$, and $\nu=1/2$ while for point C (black dot-dashed line), $\Delta_{\infty} = 0.19E_F$, $E_\text{g}^\infty=0.16 E_F$, $A = 0.2 E_F^{1+\nu}$, $\theta  = \pi/4$ and $\nu=3/4$. Inset shows the corresponding dynamics of the spin polarization, with green line for point B and red line for point C.}
\label{fig-fig6}
\end{figure}

{\bf Phase II.} In this damped phase, the magnitude of the order parameter undergoes damped oscillation and tends to a finite equilibrium value. Two examples (corresponding to points B and C in Fig.~\ref{fig-fig3}) are shown in Fig.~\ref{fig-fig6}. Here the magnitude of the order parameter can be described by the following power-law decay function \cite{Barankov06, Yuzbashyan06a, Volkov, Andreev04}
\begin{equation}
\label{cos}
|\Delta(t)| = \Delta_{\infty} + {A \over t^{\nu}} \cos(2E_\text{g}^\infty t + \theta),
\end{equation}
where $\Delta_{\infty}$ is the magnitude of the order parameter in the
asymptotic limit, which in general does not equal to the order parameter $\Delta_f$ determined by the Zeeman field $h_f$ at equilibrium. $E_\text{g}^\infty$ is the minimal band gap of the effective Hamiltonian at $t\rightarrow\infty$. It should be pointed out that, unlike the conventional BCS model,  $E_\text{g}^\infty$ does not in general equal to $\Delta_\infty$ in the current model, due to the presence of the SO coupling and the Zeeman field. The second term on the r.h.s. of Eq.~(\ref{cos}) gives rise to the decay of the order parameter. The exponent $\nu$, characterizing the power-law decay, is not a universal constant in this model. This is in distinct contrast to the conventional BCS model, where
$\nu = 1/2$  \cite{Barankov06, Yuzbashyan06a, Volkov} in the BCS limit, and $\nu = 3/2$  \cite{Andreev04} in the Bose-Einstein Condensation (BEC) limit.

In this phase, the order parameter behaves as $\Delta(t)=\Delta_\infty e^{-i2\mu_\infty t}$ in the asymptotic limit \cite{Foster13, Foster14}. Again we can gauge out the phase factor linear in time and treat $\mu_\infty$ as an effective chemical potential \cite{Foster14}. We can therefore construct a time-independent BdG Hamiltonian by replacing the chemical potential and order parameter in Eq.~(\ref{eq-Heff}) with $\mu_\infty$ and $\Delta_\infty$, respectively. This is still a dynamical phase because $\mu_{\infty} \ne \mu_{f}$, and $\Delta_{\infty} \ne \Delta_f$, where $\Delta_f$ and  $\mu_f$ are equilibrium order parameter and chemical potential with Zeeman field $h_f$. For example, for point B, we have $\Delta_f = 0.662E_F$ and $\mu_f = 0.199E_F$, whereas numerically
we obtain $\Delta_\infty \approx 0.456E_F$ and $\mu_\infty\approx-0.019E_F$. Given the asymptotic time-independent BdG Hamiltonian, the region of the dynamical topological phase can be determined by the condition
\begin{equation}
h_f^2 > |\Delta_{\infty}|^2 + \mu_{\infty}^2,
\label{eq-topo1}
\end{equation}
with a topological index $W=1$, otherwise we have a non-topological dynamical phase with $W=0$. In the following we will show that dynamical edge state can indeed be observed in the topological regime. We have also calculated the Chern number $\mathcal{C} = {i \over 2\pi} \sum_{n < 0} \int dk_x dk_y \varepsilon^{ab} \langle \partial_{k_a} \psi_{n {\bf k}} | \partial_{k_b} \psi_{n {\bf k}} \rangle$ of the system, where $n < 0$ means that we sum over all occupied bands of effective time-independent Hamiltonian [see Eq.~(\ref{eq-18}) in Methods]. We find that $\mathcal{C} = 1$ (0) in the topological (non-topological) regime defined above.

\begin{figure}
\centering
\includegraphics[width=3.4in]{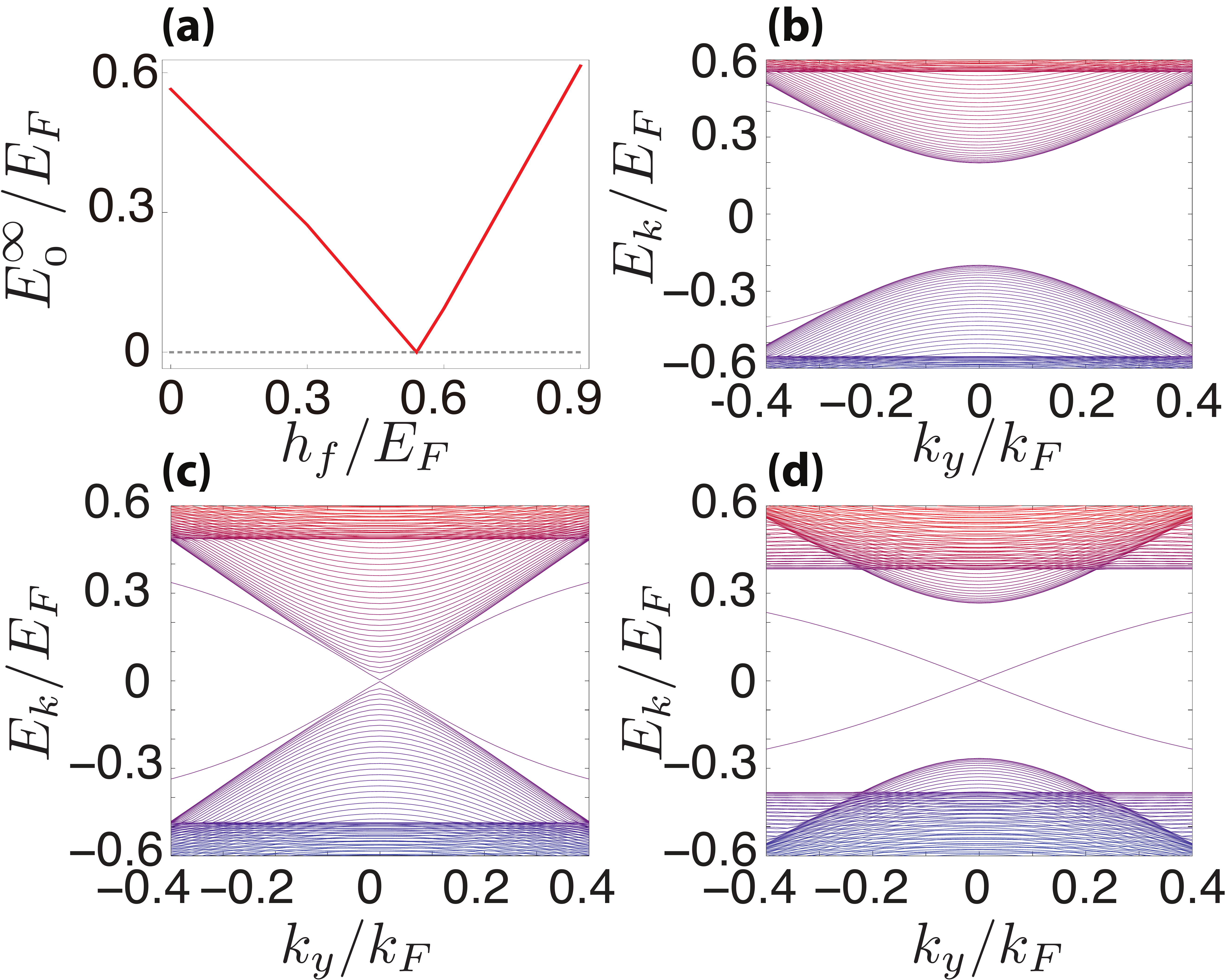}
\caption{{\bf Dynamical edge state in a strip for phase II}. (a) Energy gap at ${\bf k} = 0$, $E_0^{\infty} = |h_f - \sqrt{|\Delta_{\infty}|^2 + \mu_{\infty}^2}|$, as a function of final Zeeman field. The closing and reopening of energy gap $E_0^{\infty}$ signals a transition from a trivial phase ($W = 0$) to topological phase $W = 1$ at $h_c = 0.53E_F$. In this plot $h_i = 0.6E_F$ is used, and all the other parameters are identical to that in Fig.~\ref{fig-fig1}. (b) - (d) show the band structure in a strip geometry with hard wall boundary condition. Robust edge states with linear dispersion can be observed in the topological phase regime. The final Zeeman field from (b) to (d) are $0.4E_F$, $0.55E_F$, and $0.7E_F$, respectively.}
\label{fig-fig7}
\end{figure}

The topological nature of the system can also be manifested by examining the existence of the  Majorana edge modes. To this end, we obtain the BdG spectrum by adding a hard-wall boundary along the $x$-direction. Examples are shown in Fig.~\ref{fig-fig7}. We show the energy gap at zero momentum at $t=\infty$, $E_0^{\infty} = |h_f - \sqrt{\mu_{\infty}^2 + |\Delta_{\infty}|^2}|$, as a function of the final Zeeman field $h_f$ in Fig.~\ref{fig-fig7}(a) for fixed initial Zeeman field. The closing and reopening of the energy gap $E_0^\infty$ signals the dynamical topological phase transition. Indeed, we show that zero-energy dynamical edge state can be observed in the topological regime, see Fig.~\ref{fig-fig7}(d), where the bulk spectrum is gapped and the edge state is gapless.

\begin{figure}
\includegraphics[width=3.4in]{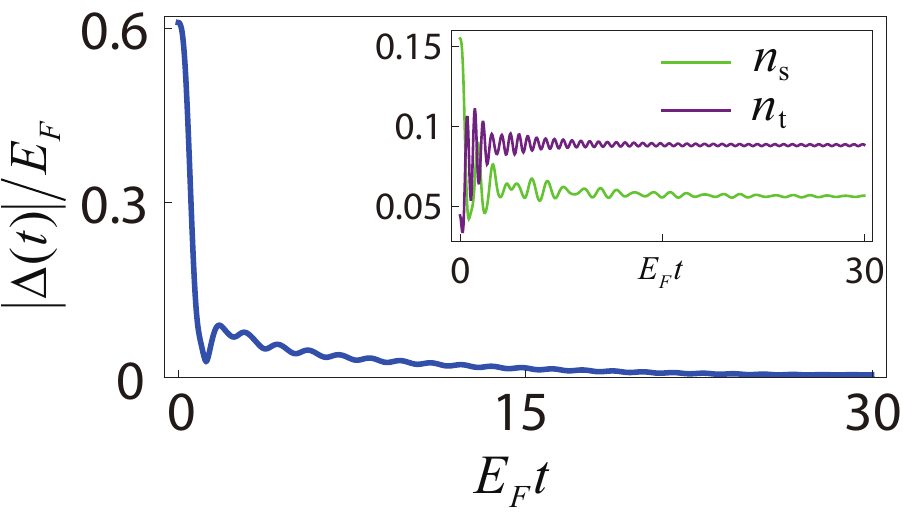}
\caption{{\bf Dynamics of order parameter and condensate fraction in phase III}. We plot the result for point D $(h_i=0.3, h_f=2.4)E_F$ in Fig.~\ref{fig-fig3}.  Inset shows the dynamics of condensate fraction of singlet pairing $n_s$ (green) and triplet pairing $n_t$ (purple), which remains finite values although $|\Delta(t)|$ approaches zero in the long-time limit.}
\label{fig-fig8}
\end{figure}

{\bf Phase III.} In phase III, the order parameter quickly decays to zero (see
Fig.~\ref{fig-fig8}) according to $\Delta(t) \sim \exp(-t/T^*)$, where $T^* \sim
1/\Delta$,  the decay time, is equal to the order parameter dynamical time (see
Discussion). However, we need to emphasize that a vanishing order parameter does
not mean that the system has become a normal gas. We demonstrate this by showing
the dynamics of both the singlet and the triplet condensate fraction, defined as
$n_s=\sum_{{\bf k}}|\langle c_{{\bf k}\uparrow}c_{-{\bf k}\downarrow}
\rangle|^2/n$ and $n_t=\sum_{{\bf k}}|\langle c_{{\bf k}\uparrow}c_{-{\bf
k}\uparrow} \rangle|^2/n$. One can see that in the long-time limit, the
condensate fraction remains non-zero even though the order parameter vanishes.
Non-zero condensate fraction means that the system still contains nontrivial pairings. 
However, the pairing field for different momenta oscillates at different frequencies, which leads to dephasing and hence a vanishing order parameter. \newline

{\Large Discussion} \newline

In this Article, we have demonstrated that dynamical topological phases can be realized in an SO coupled degenerate Fermi gas by quenching Zeeman field. The Zeeman field directly determines the topological properties of the ground state, which is completely characterized by the zero-momentum spin polarization $S_z(0)$, a quantity that is directly measurable in cold atom experiments using the standard time-of-flight technique. We have further mapped out the post-quench phase diagram according to the asymptotic behavior of the order parameter. In the undamped phase, the persistent oscillation of the order parameter may support topological Floquet state with multiple edge states. In the damped phase, the magnitude of the order parameter gradually approaches a constant via a power-law decay, and this phase contains a dynamical topological portion in certain parameter regions. One pair of edge modes can be observed in this case. In the overdamped phase, the order parameter quickly decays to zero exponentially while the condensate fraction remains finite.

The presence of the SO coupling and the Zeeman field breaks the integrability of our model. However, the same types of post-quench dynamical phases observed in our model are also present in integrable models studied previously. This raises the important question on the relationship between integrability and the long-time asymptotic post-quench behavior of the superfluid/superconducting system. This issue has been intensively investigated in other models regarding relaxation, thermalization and phase transitions \cite{Kinoshita06, Anatoli}, in which integrability plays the most essential role therein. Our work here shows that this is a rather subtle question and further studies are needed to provide a definitive answer.

We finally comment on the feasibility of observing the exotic dynamical
topological phases unveiled in this Article. The  dynamics of the superfluids
are mainly determined by two characteristic time scales, that is, the energy
relaxation time $\tau_{\epsilon} \sim E_F/E_\text{g}^2$, where $E_\text{g}$ is
the energy gap of the superfluids before quench, and the order parameter
dynamical time $\tau_{\Delta} \sim 1 /\Delta$. As we have already mentioned, the quench of the effective Zeeman field in SO coupled Fermi gases can be achieved at a time scale  much smaller than $1/E_F$ \cite{Qu13}. The
far-from-equilibrium {\it coherent} evolution can be realized when
$\tau_\epsilon > t \ge \tau_\Delta$ \cite{Barankov04, Barankov06}. The ultracold
Fermi gas provides a natural system to explore the physics in the
far-from-equilibrium condition at a time scale of $1/E_F$.  In the BCS limit
$\Delta = \sqrt{2E_FE_b}$ approaches zero. Thus $E_{\text{g}} = \Delta \ll E_F$
($\mu > 0$ in the BCS limit), and we immediately have $\tau_\epsilon \gg
\tau_\Delta$ (Using point A in Fig.~\ref{fig-fig3} as an example, we have
$\Delta \sim 0.013 E_F$, $\mu \sim 0.7 E_F$, $E_\text{g} \sim 0.009 E_F$, thus
$\tau_{\Delta} \sim 70/E_F$ and $\tau_{\epsilon} \sim 10^4/E_F \sim 160
\tau_{\Delta}$). In our system, the SO coupling and the Zeeman field can
greatly change the band structure of the superfluids. For example, the energy
gap is no longer determined solely by the order parameter and chemical
potential, but instead, it is a very complicated function of all parameters; see
Methods. At the boundary of topological phase transition, we have $E_\text{g} =
E_0 = 0$. In the vicinity of this boundary, $\tau_{\epsilon} \sim E_F / (h -
\sqrt{\Delta^2 + \mu^2})^2$, we naturally expect that $\tau_\epsilon \gg
\tau_\Delta$. We should emphasize that this condition, which can only be
realized in the BCS limit in a conventional $s$-wave superfluid, can now be realized 
very easily in a SO coupled model in the strong coupling regime because of the
different parameter dependence for these two time scales. Meanwhile, the temperature effect is also a critical issue in
ultracold atomic system. In the BCS limit,  we expect $T_c = {2E_F e^{\gamma}
\over \pi} \sqrt{E_b/2E_F} \sim {1 \over 2}\Delta$ for a conventional superfluid \cite{Bauer14}, where $\gamma
\simeq 0.577$ is the Euler's constant. The required temperature is thus very
low in order to observe the coherent dynamics of superfluids, which poses a great
challenge to the experiments \cite{Martiyanov10, Martiyanov14,
Frohlich11, Dyke2011}. This dilemma can be resolved in our model because
of the dramatic change in band structure caused by the SO coupling and the Zeeman field.
In the strong coupling regime, we expect the relevant temperature to be
determined by the Kosterlitz-Thouless transition temperature $T_\text{KT} \sim
0.1E_F$ \cite{MGong12, Bauer14}, which is experimentally accessible within current technology
\cite{Martiyanov10, Martiyanov14, Frohlich11, Dyke2011}. In fact, the
temperature effect may become important only when $T \gg \Delta$, in which case
the pairing may be destroyed.  For example, we have verified that the persistent oscillations in phase I regime can still be observed if we add a small finite temperature $T$ in the time-dependent BdG equation. More specifically, for the parameter set used in Fig.~\ref{fig-fig4}, oscillations can be observed up to $T \approx 0.2E_F$. For these reasons, 
we expect that the relevant dynamics of the order parameter and associated  {\it dynamical} topological phase transitions in phase I and 
phase II regimes can be realized using realistic cold atom setup at the currently achievable temperatures.  \newline

{\Large Methods} \newline

{\bf Equation of motion}. The basic Hamiltonian in Eq.~(\ref{eq-H1}) is solved under the Bogoliubov-de Gennes formalism by defining the order parameter
$\Delta = -g \sum_{{\bf k}} \langle c_{{\bf k}\uparrow} c_{-{\bf k}\downarrow} \rangle$. Then the dynamics of the model is governed by
the effective Hamiltonian $\mathcal{M}_{{\bf k}}$ in Eq.~(\ref{eq-Heff}). The initial value of chemical potential and order parameter are determined by minimizing the thermodynamical potential \cite{MGong11, MGong12, Sato09,ZhouJ11,Seo12,Iskin11,HuiH11,Pu2, Pu9, CQu13,WZhang13,Liu13,CC13}, which is equivalent to solving the gap equation
\begin{equation}
\label{gap}
\sum_{\bf k} \left[\frac{1}{2\epsilon_{\bf k}+E_b} - \sum_{s =\pm} \frac{E_0+ sh^2}{4E_0E_{{\bf k}s}^+}\right] = 0,
\end{equation}
and number equation
\begin{equation}
\label{num}
n=\sum_{\bf k}\left[1- \sum_{s=\pm} \frac{\xi_{\bf
k}}{2}\frac{E_0+s(h^2+\alpha^2k^2)}{E_0E_{{\bf k}s}^+}\right],
\end{equation}
where $E_{{\bf k}\pm}^\eta=\eta\sqrt{\xi_{\bf k}^2+\alpha^2k^2+h^2+|\Delta|^2\pm2E_0}$
are the excitation energies, with $\eta=\pm$ correspond to the particle and hole branches respectively and $E_0=\sqrt{h^2(\xi_{\bf k}^2+|\Delta|^2)+\alpha^2k^2\xi^2_{\bf k}}$.
Throughout our numerical calculations, the energy unit is chosen as the Fermi energy $E_F=k^2_F/2m$, where $k_F=\sqrt{2\pi n}$ is the Fermi wavenumber for a
non-interacting Fermi gas without SO coupling and Zeeman field in 2D. We only consider the physics at zero temperature. Throughout this work, we choose the binding energy $E_b = 0.2E_F$, and the corresponding scattering length $k_Fa_{\text{2D}} = \sqrt{2E_F/E_b} \simeq 3.1623$ and $\ln(k_Fa_{\text{2D}}) \simeq 1.1513$, which is around the
BEC-BCS crossover regime. Since $k_Fa_{\text{2D}} > 1$, the superfluid is
composed of weakly bound Cooper pairs. For a typical Fermi gas of $^6$Li, $k_F
\sim 1/\mu$m, $E_F \sim 1 \text{kHz}$ , and the relevant time scale discussed in this work is $1/E_F \sim 1 \text{ms}$ . The long time collective oscillations of degenerate Fermi gas
that much longer than this relevant time scale has been demonstrated in experiments \cite{Bertaina11, Altmeyer07}.

The initial value of $\mu$ and $\Delta$ can be directly determined by solving the gap equation and number equation self-consistently, with which the
initial wavefunction at $t = 0^-$ is constructed. In this case, the topological phase can be realized when \cite{MGong11, MGong12},
\begin{equation}
h^2 > \mu^2 + \Delta^2.
\label{eq-topo0}
\end{equation}
However, this phase can only be realized in the BEC-BCS crossover regime. A simple but intuitive argument is that, in both the BEC and BCS limit,
$\mu^2 \gg \Delta^2$, thus in these two limits, $|h| \gg |\Delta|$, and the
pairing can be easily destroyed by the Pauli depairing effect. As a result,
the topological phase can only be realized in a small parameter window near the
strong coupling regime. In the following, we choose Zeeman field as the quench parameter instead of others, because it is the most easily controllable parameter in the current experiments  \cite{Lin11,Zhang12,Wang12,Cheuk12, Qu13, Chris14}.
Moreover, the Zeeman field is directly relevant to the topological boundary while many-body interaction is not; see Eq.~(\ref{eq-topo1}) and
Eq.~(\ref{eq-topo0}).

Immediately after the quench of the Zeeman field, the system's wavefunction is assumed to keep the BCS form
\begin{equation}
|\Psi(t)\rangle = \prod_{{\bf k},\pm} f^\dag_{{\bf k}\pm} \Psi_{{\bf k}} |0\rangle,
\end{equation}
where $f_{{\bf k}\pm}$ satisfy the time-dependent BdG equation in
Eq.~(\ref{eq-tBdG}). In fact Eq.~(\ref{eq-tBdG}) is equivalent to
\begin{equation}
i\partial_t |\Psi(t)\rangle = H |\Psi(t)\rangle.
\end{equation}
The above semiclassical equation can be derived from $\delta_{\Psi^*} \mathcal{L} = 0$, with $\mathcal{L} = \langle \Psi|i\partial_t - H| \Psi\rangle$.

{\bf Spin texture at zero momentum}.
The Hamiltonian at ${\bf k} = 0$ can be reduced to
\begin{equation}
\mathcal{M}_{\bf{k}=0}=\left(
\begin{array}{cccc}
-\mu+h       & 0 & \Delta               & 0 \\
0 & -\mu-h  & 0                    & \Delta \\
\Delta^*         &   0            &\mu+h       & 0 \\
0                   & \Delta^*    & 0 & \mu-h
\end{array}
\right).
\end{equation}
We first consider the spin texture in the stationary condition. If $|h|<\sqrt{\mu^2+\Delta^2}$, the two eigenvectors with positive eigenvalues are $f_{0,+}=(\frac{-\mu+\lambda}{\sqrt{\Delta^2+(\mu-\lambda)^2}},0,\frac{\Delta}{\sqrt{\Delta^2+(\mu-\lambda)^2}},0)^T$ and $f_{0,-}=(0,\frac{-\mu+\lambda}{\sqrt{\Delta^2+(\mu-\lambda)^2}},0,\frac{\Delta}{\sqrt{\Delta^2+(\mu-\lambda)^2}})^T$ respectively, where $\lambda=\sqrt{\mu^2+\Delta^2}$. Then, from the expression of $\langle S_z({\bf k})\rangle$ in the main text, we can directly get  that $\langle S_z(0)\rangle=0$ and $Q=0$, which means the system has topologically trivial spin texture. In contrast, if $|h|>\sqrt{\mu^2+\Delta^2}$, the two eigenvectors will become $f_{0,+}=(\frac{-\mu+\lambda}{\sqrt{\Delta^2+(\mu-\lambda)^2}},0,\frac{\Delta}{\sqrt{\Delta^2+(\mu-\lambda)^2}},0)^T$ and $f_{0,-}=(\frac{-\mu-\lambda}{\sqrt{\Delta^2+(\mu+\lambda)^2}},0,\frac{\Delta}{\sqrt{\Delta^2+(\mu+\lambda)^2}},0)^T$ respectively. So, in this region, we will have $\langle S_z(0)\rangle=-1$ and $Q=-1$, which means the system has a topologically nontrivial spin texture.  We see that the sudden changes of $S_z(0)$ is due to the band inversion transition across the critical point.

The time-evolution of $Q$ can be discussed in a similar way. When ${\bf k} = 0$,
we have $i\partial_t q_{\pm}=\Delta^* v_{\pm}-hq_{\pm}$ and $i\partial_t
p_{\pm}=\Delta^* u_{\pm}+hq_{\pm}$. Substituting these equations into
$\dot{S}_z(t) = \sum_{\eta=\pm} \dot{q}^*_\eta q_\eta + q^*_\eta\dot{q}_\eta -
\dot{p}^*_\eta p_\eta + p^*_\eta\dot{p}_\eta$, we immediately find that
$\dot{S}_z(t) = 0$. Thus we have the important conclusion that $\langle
S_z(0,t)\rangle=\langle S_z(0,0)\rangle$, which means $Q$ remains unchanged over
time. We need to emphasize that this spin texture is totally different from the
pseudospin texture discussed in Ref.  \onlinecite{Foster13}. The {\it true} spin
texture discussed in this work can be directly probed in experiments from the
time-of-flight imaging \cite{Alba11,Cheuk12, Sau11} in which $Q(t) = Q(0)$ can be 
directly verified.

{\bf Dynamical edge state in phase II regime}. In this regime, the magnitude of the order parameter will gradually approach a constant, while its phase factor oscillates
periodically, i.e., $\Delta(t)\rightarrow\Delta_{\infty} e^{-i\mu_{\infty}t }$ up to a trivial constant, which is the only time-dependent parameter in the BdG Eq.~(\ref{eq-tBdG}). This oscillating phase can be gauged out by defining $f_{\bf k \pm} = (\tilde{u}_{{\bf k}\pm}e^{-i\mu_\infty t},\tilde{v}_{{\bf k}\pm}e^{-i\mu_\infty t},\tilde{p}_{{\bf k}\pm}e^{+i\mu_\infty t},\tilde{q}_{{\bf k}\pm}e^{+i\mu_\infty t})^T e^{-iE_{{\bf k,\pm}}t}$, where $\tilde{f}_{{\bf k}\pm} = (\tilde{u}_{{\bf k}\pm},\tilde{v}_{{\bf k}\pm},\tilde{p}_{{\bf k}\pm},\tilde{q}_{{\bf k}\pm})^T$. Inserting this wavefunction to Eq.~(\ref{eq-tBdG}), we find that
\begin{equation}
 \tilde{\mathcal{M}}_{{\bf k }}\tilde{f}_{{\bf k}\pm} =   E_{{\bf k,\pm}}\tilde{f}_{{\bf k}\pm},
\label{eq-18}
\end{equation}
where $\tilde{\mathcal{M}}_{{\bf k }}$ is identical to Eq.~(\ref{eq-Heff})
except that $\mu = \mu_{\infty}$ and $\Delta = \Delta_{\infty}$. We immediately
see that $\mu_{\infty}$ is the effective chemical potential of the model in the
quasi-equilibrium condition. Note that this phase is still dynamical phase
because the $\mu_{\infty} \ne \mu_f$ and $\Delta_{\infty} \ne \Delta_f$, where
$\mu_f$ and $\Delta_f$ are equilibrium chemical potential and order parameter
with Zeeman field $h_f$; see our numerical results in the main text. We did not observed the abrupt change of order parameter in all our calculations \cite{tbdgXianlong}.

This model can support dynamical edge state in the topological regime defined by Eq.~(\ref{eq-topo1}). Similar to the analysis in Ref.~[\onlinecite{MGong11}], we can prove exactly that the bulk system is always fully gapped except at the critical point of $h_c$ for ${\bf k} = 0$. Thus the closing and reopening of the gap provides important indications for topological phase transition. To see the topological phase transition more clearly, we consider a strip superfluids with length $L$ by imposing hard wall boundary at the $x$ direction. To this end, we replace $k_x\rightarrow -i\partial_x$, while $k_y$ remains as a good quantum number. Along the $x$ direction, we construct the wavefunction using plane wave basis \cite{Chan14, HuiH14}
\begin{equation}
\left(\begin{array}{c}
\tilde{u}(x)\\
\tilde{v}(x)\\
\tilde{p}(x)\\
\tilde{q}(x)\end{array}\right)=\sum_{n=1}^{N_{\text{max}}}\sqrt{\frac{2}{L}}\sin\left(\frac{n\pi x}{L}\right)\left(\begin{array}{c}
\tilde{u}_{n}\\
\tilde{v}_{n}\\
\tilde{p}_{n}\\
\tilde{q}_{n}\end{array}\right),
\end{equation}
where $N_{\text{max}}$ is the basis cutoff. Upon inserting this ansatz into Eq.~(\ref{eq-18}), we can convert the matrix $\tilde{\mathcal{M}}_{\bf k}$ into a $4N_{\text{max}}$ by $4N_{\text{max}}$ matrix, whose diagonalization directly leads to the protected modes of dynamical edge state dictated by nontrivial topological invariants. Empirically, we found $N_{\text{max}}=200$ is a good basis cutoff for a long strip with $L=200 k_F^{-1}$. The numerical results are presented in Fig.~\ref{fig-fig7}.

{\bf Dynamical Floquet state in phase I regime}. For a general periodically
driven Hamiltonian $\hat{H}_{\textbf{k}}(t)$ with period $T$, that is,
$\hat{H}_{\textbf{k}}(t) = \hat{H}_{\textbf{k}}(t+T)$, we invoke the Floquet
theorem to examine its Floquet spectrum. We can define the Floquet
wavefunction as $|\varphi_{{\textbf{k}}}(t)\rangle = e^{-i
\varepsilon_{{\textbf{k}}}t} |\Psi_{{\textbf{k}}}(t)\rangle$, where $\varepsilon$
is the quasiparticle spectrum and $|\Psi_{{\textbf{k}}}(t)\rangle =
|\Psi_{{\textbf{k}}}(t+T)\rangle$, i.e., the wavefunction in the time-domain is
also a periodic function. We are able to expand $|\Psi_{{\textbf{k}}}(t)\rangle
= \sum_n \Psi_{n,{\textbf{k}}}e^{in2\pi t/T}$. Then we can map the
time-dependent model to the time-independent model $\mathcal{H}_{4(2N + 1)\times
4(2N+1)} \Psi_\textbf{k} = \varepsilon \Psi_\textbf{k}$, where  $\Psi_\textbf{k} = (\Psi_{-N,{\textbf{k}}},
\Psi_{-N+1,{\textbf{k}}},\cdots, \Psi_{N-1,{\textbf{k}}},
\Psi_{N,{\textbf{k}}})^T$ with $N$ being a truncation of frequencies. This model has
particle-hole symmetry defined as $\Sigma = \mathbb{I}_{2N+1}\otimes\sigma_x K$
with $K$ being the conjugate operator and $\mathbb{I}_{2N+1}$ being an identity matrix. In our two dimensional model, this equivalent model belongs to topological $D$ class with index $\mathbb{Z}$.

In the long-time limit, the order parameter in phase I approaches $\Delta(t)=|\Delta_{\infty}(t)|e^{-2i\mu_{\infty}t+i\varphi(t)}$, where $|\Delta_{\infty}(t)|$ is periodic in time, e.g. see Fig.~\ref{fig-fig4}.
We make a gauge transformation, similar to that in Eq.~(\ref{eq-18}), by identifying $\mu_\infty$ as the effective chemical potential,
and we obtain $\tilde{\mathcal{M}}_{\bf k}(t)$ from Eq.~(\ref{eq-Heff}) by replacing $\mu$ with $\mu_{\infty}$ and $\Delta$ with $|\Delta(t)|e^{i\varphi(t)}$. Obviously, $\tilde{\mathcal{M}}_{\bf k}(t) = \tilde{\mathcal{M}}_{\bf k}(t+T)$,
where $T$ is the period determined by both $|\Delta(t)|$ and $\varphi(t)$.  Now we assume the eigenvectors of the above effective Hamiltonian to be
$\tilde{f}_{\bf k\pm}(t)=\Phi_{\bf k\pm}(t) e^{-i\varepsilon_{\text{\bf k}\pm} t}$, where $\Phi_{\bf k\pm}(t+T)=\Phi_{\bf k\pm}(t)$. Then we have
\begin{equation}
\left(\tilde{\mathcal{M}}_{\bf k}(t)-i\partial_{t}\right)\Phi_{\bf k\pm}(t)=\varepsilon_{\text{\bf k}\pm}\Phi_{\bf k\pm}(t),
\label{phaseIbdg}
\end{equation}
where $\varepsilon_{\text{\bf k}\pm}$ is the quasiparticle dispersion. Similar to the discussion of dynamical edge state in the previous section, we impose
a hard wall boundary condition along $x$ direction with length $L$. We expand the wavefunction in the following way,
\begin{equation}
		\Phi_{\bf k\pm} = \mathcal{A} \sum_{n = 1}^{N_{\text{max}}} \sum_{m =-M_{\text{max}}} ^{M_{\text{max}}} \sin(\frac{n\pi x}{L})e^{\frac{2im\pi t}{T}}
\left(\begin{array}{c}
		\tilde{u}_{nm}^{\pm}\\
		\tilde{v}_{nm}^{\pm}\\
		\tilde{p}_{nm}^{\pm}\\
		\tilde{q}_{nm}^{\pm}
\end{array}\right),
\end{equation}
where $N_{\text{max}}$ and $M_{\text{max}}$ are basis cutoff for the spatial and temporal expansion and $\mathcal{A} = (2/LT)^{1/2}$. Then Eq.~(\ref{phaseIbdg}) can be recast into a sparse self-adjoint complex matrix form of size $(4N_{\text{max}}\times(2M_{\text{max}}+1))\times(4N_{\text{max}}\times(2M_{\text{max}}+1))$. The direct diagonalization of this matrix gives rise to Floquet spectrum. In practice, since we are only concerned with the eigenenergies close to zero, we could utilize the shift and invert spectral transformation and compute   only a portion of eigenenergies using the ARPACK library routines. For instance, we choose cutoffs as $N_{\text{max}}=200$ and $M_{\text{max}}=15$ and only compute 500 eigenvalues around zero-energy out of total 24800 ones for a given $k_y$. The results are presented in Fig.~\ref{fig-fig5}. The robustness of these protected edge states are also examined by slightly changing the model parameters, in which we find that the linear dispersions of these edge states are unchanged.

{\bf Majorana fermions at zero momentum}. The intrinsic particle-hole
symmetry ensures that the zero energy states are Majorana fermions at $k_y = 0$
in a strip along $y$ direction. The wavefunction of these Majorana fermions can be constructed
using $\psi = \psi_i \pm \Sigma \psi_i$, where $\psi_i$ is the quasiparticle wavefunctions 
with energy approach zero. We can observe: (1) The edge states have well-defined chirality which
are defined as $\Sigma \psi = \pm \psi$; (2) These edge modes are well-localized
at the boundaries when the strip width is much larger than the coherent length;
(3) All the edge states with the same chirality are localized at the same boundary;
(4) For a general random potential $\mathcal{V}$, the matrix elements $\langle \psi_i | \mathcal{V} | \psi_j\rangle \equiv 0$
when $\psi_{i}$ and $\psi_j$ have the same chirality. Thus these zero energy states are robust against
perturbations and the Majorana edge states can always be observed in the topological phase regime. We have numerically
verified this point by slightly varying the parameters of the Hamiltonian and we
find that the edge modes will not be gapped out without closing the quasiparticle
energy gap at zero momentum.

{\bf Acknowledgements} We acknowledge valuable discussions with Matthew S. Foster, Yuxin Zhao and Jia Liu. H.P. thanks the hospitality of KITPC where part of the writing is done, and helpful discussions with Victor Gurarie. Y. D. acknowledges financial support from NSFC (Grant No. 11304072,  11205043 and 11274085), and the fund of Hangzhou-City Quantum Information and Quantum Optics Innovation Research Team. M.G. is supported by Hong Kong RGC/GRF Projects (No. 401011 and No. 2130352), University Research Grant (No. 4053072) and The Chinese University of Hong Kong (CUHK) Focused Investments Scheme. L. D. and H. P. are supported by the ARO Grant No. W911NF-07-1-0464 with the funds from the DARPA OLE Program, the Welch foundation (C-1669) and the NSF.

{\textbf{Author Contributions}} Y.D. and L.D. have equal contributions to this work. All authors contributed to writing and revising the manuscript, and participated in discussions about this work. Correspondence and requests for materials should be addressed to M.G. (skylark.gong@gmail.com) and/or H.P. (hpu@rice.edu).

{\textbf{Competing Interests}}  The authors declare no competing financial interests.

\end{document}